\documentclass[runningheads]{llncs}
\usepackage{dsfont}
\usepackage{graphicx}
\usepackage{xcolor}
\usepackage{listings}
\usepackage{amssymb, amsmath}
\usepackage{subcaption}
\usepackage{url}
\begin{document}
\title{Distributed Objective Function Evaluation for Optimization of Radiation Therapy Treatment Plans}
\titlerunning{Distributed Optimization for Radiation Therapy}
%
\author{Felix Liu \inst{1,2} \and
Måns I. Andersson \inst{1} \and
Albin Fredriksson \inst{2}\and
Stefano Markidis \inst{1}}
\authorrunning{F. Liu et al.}
%
\institute{KTH Royal Institute of Technology \and
RaySearch Laboratories}
\maketitle              
\begin{abstract}
The modern workflow for radiation therapy treatment planning involves mathematical optimization to determine optimal treatment machine parameters for each patient case. The optimization problems can be computationally expensive, requiring iterative optimization algorithms to solve. In this work, we investigate a method for distributing the calculation of objective functions and gradients for radiation therapy optimization problems across computational nodes. We test our approach on the TROTS dataset--- which consists of optimization problems from real clinical patient cases---using the IPOPT optimization solver in a leader/follower type approach for parallelization. We show that our approach can utilize multiple computational nodes efficiently, with a speedup of approximately 2-3.5 times compared to the serial version.
\end{abstract}
\section{Introduction}
Radiation therapy is one of the most common forms of cancer treatment today. Before a patient undergoes treatment, the control parameters for the treatment machine, such as the dose rate and the shape of the aperture through which to irradiate the patient, must be determined. This process is known as \emph{treatment planning}. Ultimately, the goal of the treatment is to deliver sufficient dose to the tumor to kill cancerous cells, while sparing surrounding healthy tissue.
\\ \\
In modern radiation treatment planning, mathematical optimization is used to determine parameters for the treatment machine (often a linear- or particle accelerator). The planning process typically begins after a CT scan of the patient has been imported into a \emph{treatment planning system} (TPS), a software product designed for radiation treatment planning. Contours of important structures such as risk organs and the tumor---the regions-of-interest (ROIs)---in the patient are drawn on the CT. Optimization functions in the form of objectives and constraints corresponding to desirable qualities of the dose delivered to the ROIs are defined, yielding a mathematical optimization problem. The optimization problem is then solved numerically, and, if needed, the process is repeated in a trial-and-error fashion where the optimization functions and possible importance weights are changed until a high-quality plan is achieved.
\\ \\
The treatment planning process is time-consuming, in part because the optimization problem is computationally demanding. Efficient algorithms for solving the problem is crucial, both for efficiency at the clinics and for the quality of the resulting treatment plans. Due to the high computational demand, utilizing HPC resources such as accelerators or distributed computing clusters is an important step.
\\ \\
In this work, we propose a method for distributing the computation of objective function, constraints and gradients---in many cases an important computational bottleneck---across multiple computational nodes. We show the effectiveness of our approach by utilizing the IPOPT solver \cite{wachter2006implementation}, a general software library for nonlinear optimization, together with our method for distributing optimization function evaluations across nodes, on optimization problems from radiation therapy. We study problems from the TROTS dataset, an open dataset with data for optimization problems from real patients treated for cancers in the head-and-neck region. We provide an implementation to calculate function values and gradients using input data from TROTS in a distributed fashion and show that our approach can produce solutions of high quality while being able to effectively utilize distributed computing resources, with approximately a 2-3x speedup compared to the single node version.

\section{Background and Related Work}
We consider treatment planning for high-energy photons using the treatment technique \emph{volumetric modulated arc therapy} (VMAT) \cite{otto2008volumetric}, where the gantry head of the treatment machine is continuously rotated around the patient during the treatment. The optimization variables in our problems are \emph{beamlet weights}, which are intensity values for the delivered beams in the plane in front of the gantry. From the beamlet weights, the actual treatment machine settings required to deliver such an intensity profile at each beam angle can be determined and a deliverable plan can be created.
\subsection{TROTS Dataset}
The TROTS dataset \cite{breedveld2017data} consists of data for patients with cancers in the head-and-neck region, liver or prostate. The dataset contains objective functions and constraints for the dose in each ROI of the patient, which form the nonlinear optimization problem to be solved. Furthermore, \emph{dose influence matrices} are provided for each ROI separately. The dose influence matrix gives the relation between the optimization variables, beamlet weights, and the resulting dose in the ROI.
\\ \\
We note that having separate dose matrices for each ROI is not ubiquitous in radiation therapy optimization, it is also possible to provide a single dose matrix covering the entire relevant part of the patient volume and to extract the doses for each ROI from the total dose. Since there is only a single dose matrix in this case, distributing the computation becomes more complicated, in which case it may be more natural to look to GPU accelerators instead, see for instance the work in \cite{liu2021accelerating}.
\begin{table}[]
    \centering
    \begin{tabular}{lll}
       Name & $f(x)$ & Comment \\
       \hline LTCP: &  $ \frac{1}{n} \sum_{i=1}^n e^{-\alpha (d_i(x) - \hat{d})} $ & where $\alpha$ is a scalar  \\
       min/max dose: & $g(d(x))$ & g is min/max \\
       Mean dose & $ \frac{1}{n} \sum_{i=1}^n d_i(x)$ & \\
       Generalized mean: & $\left(\frac{1}{n} \sum_{i=1}^n d_i(x)^a\right)^{\frac{1}{a}}$ & a is a parameter \\
       Quadratic: &  $ \frac{1}{2} x^T A x + b^T x + c$ & a,b are constant vectors
    \end{tabular}
    \vspace{0.2cm}
    \caption{The different types of optimization functions used in the TROTS dataset problems we have used. In the following $f(x)$ denotes an optimization function, $d_i(x)$ is the dose in voxel $i$ of a given ROI, and $\hat{d}$ denotes a desired dose level.}
    \label{tab:my_label}
\end{table}
For our experiments, we use the same optimization functions as provided in TROTS, with the exception of the min and max dose constraints, which we substitute with quadratic penalties of the form:
\begin{equation*}
    f(x) = \frac{1}{n} \sum_{i=1}^n (g(d_i(x) - \hat{d}, 0))^2
\end{equation*}
where $g$ again is either $\min$ or $\max$.
\\ \\
The dependence between the dose $d(x)$ for a given ROI and the optimization variables $x$ is linear. To calculate the dose for a given ROI and value of $x$, we simply multiply $x$ by the dose influence matrix $A$ for that ROI: $d(x) = Ax$. The TROTS dataset provides one dose influence matrix for each ROI, which is stored as a sparse matrix.
\\ \\
The optimization problem for a TROTS case is then specified using a weighted sum of the objectives, shown in Table~\ref{tab:my_label} on the different ROIs, together with the constraints on ROIs in the following form:
\begin{align*}
    \min_{x} \quad &\sum_{i=1}^n w_i f_i(x) \\
    \text{s.t.} \quad &g_i(x) \leq 0 \\
                           &x \geq 0
\end{align*}
Here, $f_i(x)$ are the objectives for the different ROIs, with $w_i$ being their corresponding weight and $g_i(x)$ are the constraints on doses in the ROIs.
\subsection{Dose Influence Matrices}
\begin{figure}[h!]
    \centering
    \begin{subfigure}[b]{.55\textwidth}
        \centering
        \includegraphics[width=\textwidth]{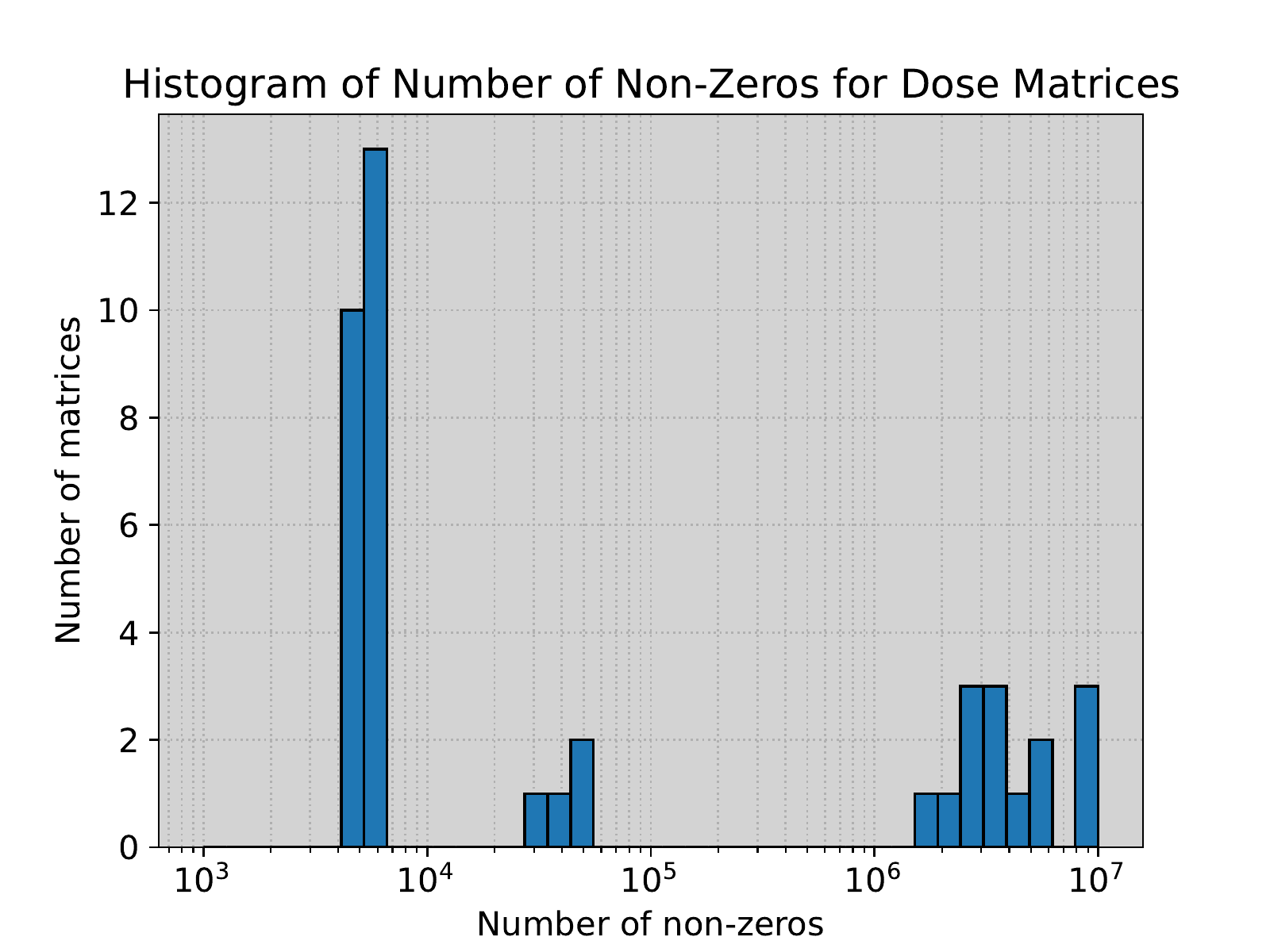}
        \caption{Histogram of the distribution of the number of non-zeros for the dose influence matrices in the Head-and-Neck problem 01 from the TROTS dataset.}
        \label{fig:nnz_hist}
    \end{subfigure}
    \begin{subfigure}[b]{.47\textwidth}
        \centering
        \includegraphics[width=\textwidth]{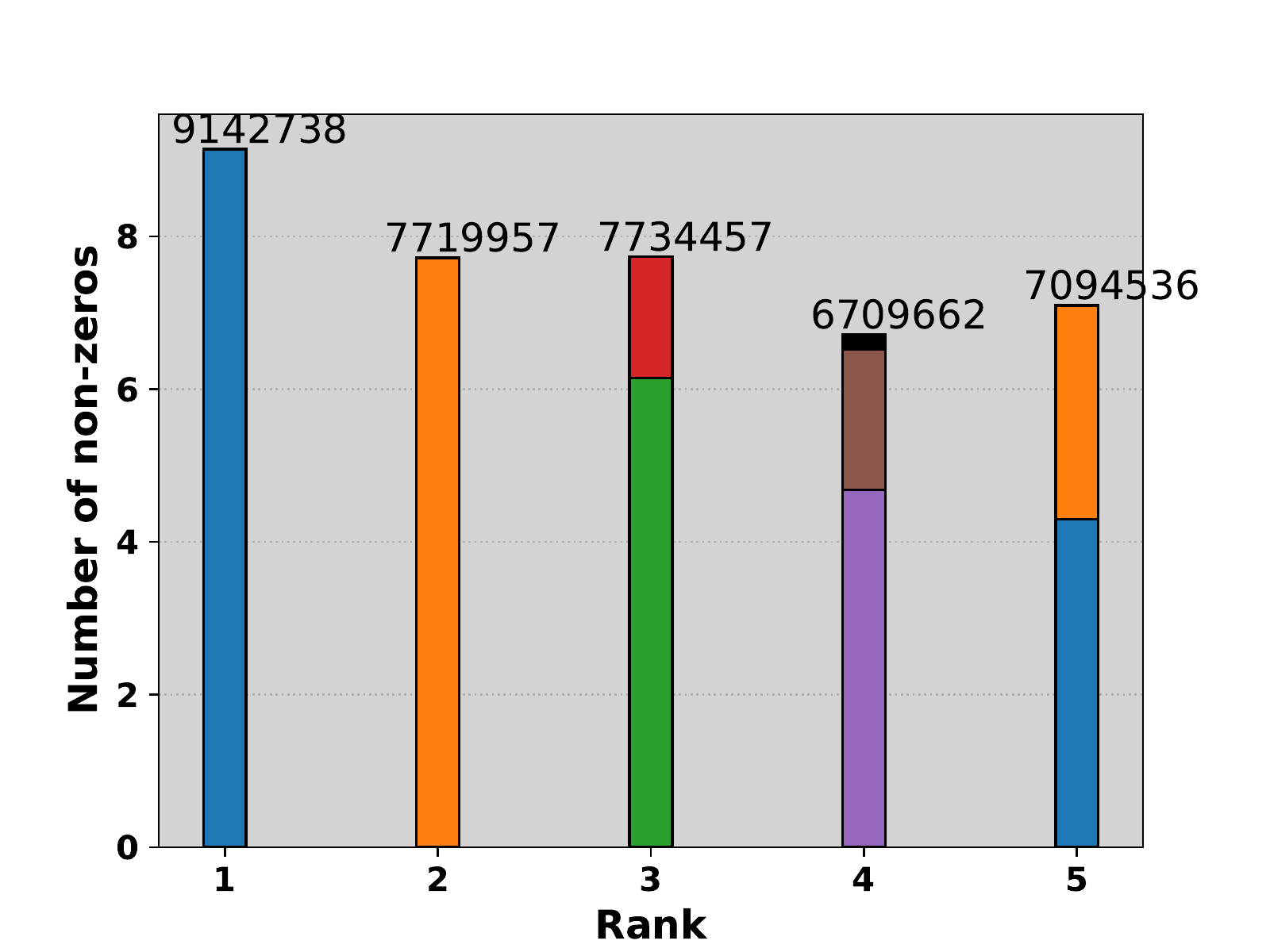}
        \caption{Greedy distribution of the dose influence matrices for objective functions.}
        \label{fig:obj_nnz_distrib}
    \end{subfigure}
    \hfill
    \begin{subfigure}[b]{.47\textwidth}
        \centering
        \includegraphics[width=\textwidth]{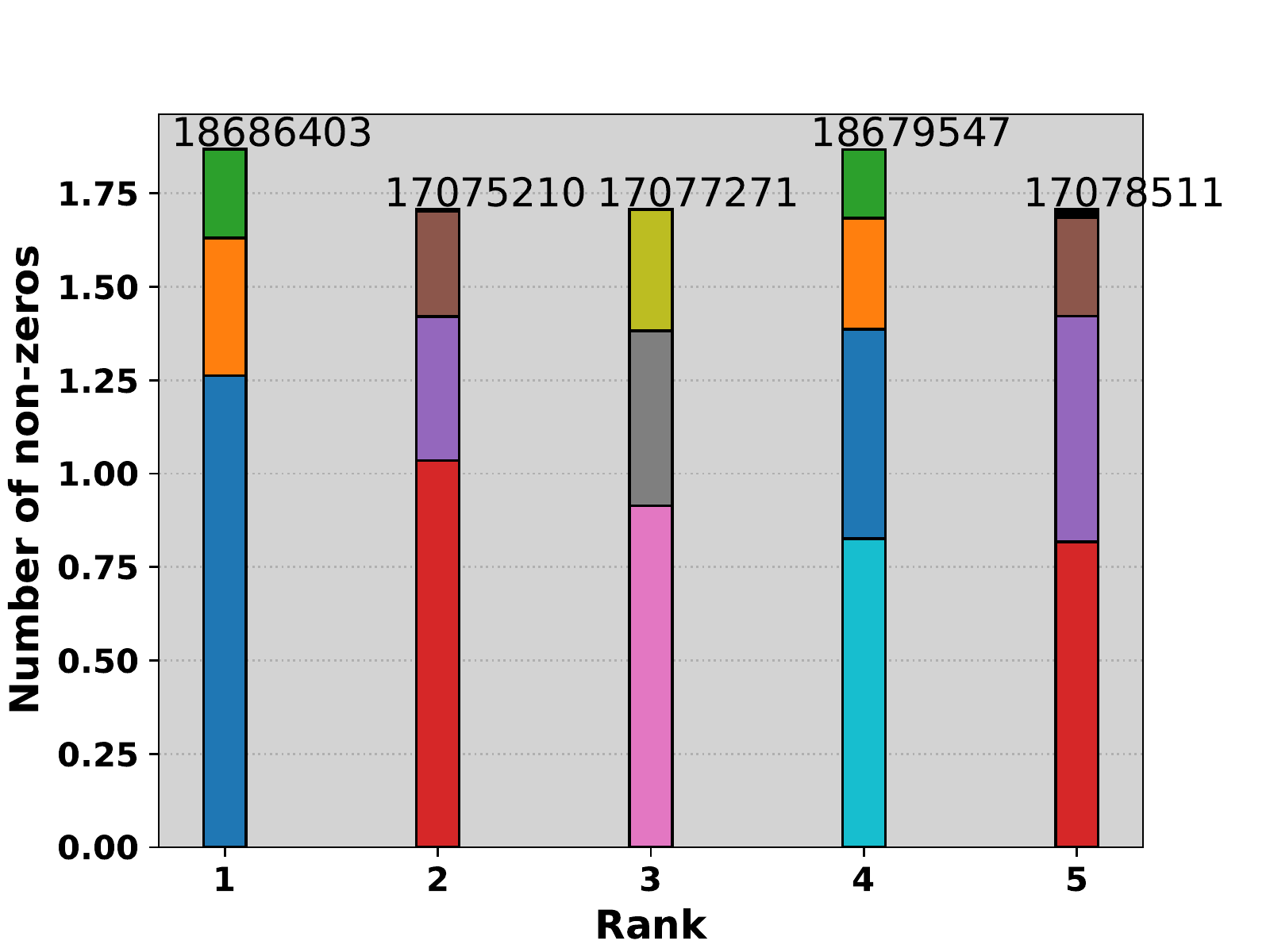}
        \caption{Greedy distribution of the dose influence matrices for constraints.}
        \label{fig:cons_nnz_distrib}
    \end{subfigure}
    \caption{Distribution of the dose influence matrices using the greedy algorithms on six ranks (one rank reserved for IPOPT). The histogram of the distribution of non-zeros of the dose matrices is shown on top. The bottom figure shows the distribution of dose matrices produced by our greedy heuristic with MPI ranks on the x-axis and corresponding colored bars being the sizes of the dose matrices assigned to that rank. The total size of all dose matrices for each ranks is shown above each bar.}
    \label{fig:nnz_distrib}
\end{figure}

Since all the optimization functions used in the optimization problems are functions of dose in the different regions of interest in the patient, dose calculation is an important computational kernel. The dose calculation in our case is a sparse matrix-vector product, $d = Ax$, with the dose influence matrices $A$ being provided by the TROTS dataset. Note also that in the case of TROTS, the dose influence matrices are given for each ROI separately, instead of for the patient volume as a whole.
\\ \\
The size of the dose influence matrix varies quite significantly between the ROIs, meaning that the computational time needed to compute the optimization function for the different ROIs varies significantly. A histogram of the number of non-zero values in the different dose influence matrices (there are approximately 40 in total) is shown in Fig.~\ref{fig:nnz_hist}. We see that the number of non-zeros varies significantly, with some matrices having approximately 5000 non-zero elements, while others have closer to 10 million. 
\subsection{Radiatiation Therapy Plan Quality} \label{sec:DVH}
The ultimate goal when solving optimization problems in radiation therapy treatment planning is to create treatment plans of high quality. To this end, the mathematical optimization problem can be seen as a proxy, providing a way to produce plans with desirable dose characteristics. Considering this, the quality of the resulting plan should not only be evaluated based on how accurately the optimization problem was solved, but also using other dose-based metrics which may reflect plan quality more accurately.
\\ \\
One common method for evaluating and comparing treatment plans is using dose-volume histograms (DVH) \cite{drzymala1991dose}. DVHs are defined using the so called volume-at-dose metric. For a given ROI with $n$ voxels (with identical sizes), discretized dose $d \in \mathcal{R}^n$ and dose level $\hat{d}$, the volume-at-dose $V_{\hat{d}}(d)$ is defined as:
\begin{equation*}
    V_{\hat{d}}(d) = \frac{\sum_{i=1}^{n} \mathds{1}(d_i \geq \hat{d})}{n},
\end{equation*}
where $\mathds{1}$ is the indicator function. Informally, the volume-at-dose gives the proportion of the ROI volume receiving a dose of at least $\hat{d}$. A DVH for a given ROI is a plot showing $V_{\hat{d}}$ as a function of different dose levels $\hat{d}$, and can be used to visually compare dose distributions from different plans.
\section{Methodology}
\subsection{Serial Version and Data Preprocessing}
While the TROTS dataset provides the data required to specify the optimization problems for each of its patient cases, it does not provide the code to compute the objective functions, constraints or gradients, or code to interface the data to optimization libraries. To be able to interface the problem to the IPOPT solver, we have developed a C++-library (which is available on Github\footnote{\url{https://github.com/felliu/Optimization-Benchmarks}}) to enable the use of general optimization libraries on the dataset. The library provides a \texttt{TROTSProblem} class, which represents a single TROTS optimization problem and provides member functions to compute objective functions, constraints and gradients. Finally, to interface the TROTS problem to the IPOPT optimization library, we simply use IPOPT's C++-interface. Note also that our library provides functions to compute function values and first-derivatives only, meaning that the optimization solver used needs to be able to run using only first-derviatives. This does not exclude the use of second-order methods which incorporate Hessian information in the optimization however, since the Hessian can be approximated using quasi-Newton methods \cite{dennis1977quasi}, which are supported in IPOPT.
\\ \\
IPOPT supports the use of multiple different linear solvers to solve the linear systems arising internally from its optimization algorithm. In general, one can expect that the overall performance of the optimization solver depends on the choice of linear solver. Initially, we tried IPOPT using the linear solvers MUMPS, and MKL Pardiso (a part of Intel MKL), since those packages are freely available. We used IPOPT's internal timers to compare the two linear solvers, and the results are summarized in Table~\ref{tab:solver_comp}. The optimization was run for a total of 3000 iterations.
\begin{table}[h!]
    \centering
    \begin{tabular}{c|c|c|c}
        Setup & Function Evaluation (s) & PD System Solution (s) & Total Time (s) \\
        \hline
        IPOPT w. MUMPS & 248.607 & 598.884 & 891.163 \\
        IPOPT w. MKL Pardiso & 212.223 & 58.951 & 279.399 \\
        \hline
    \end{tabular}
    \caption{Timings comparing IPOPT using MUMPS and MKL Pardiso. All times are wall-clock times and measured in seconds.}
    \label{tab:solver_comp}
\end{table}
As seen in Table~\ref{tab:solver_comp}, we get significantly better performance when using the MKL Pardiso linear solver, compared to MUMPS. Thus we use the MKL Pardiso solver for the remainder of this work. 
\\ \\
When using IPOPT with MKL Pardiso, we see that the computational time becomes dominated by the function evaluations, taking approximately 76\% of the total wall clock time in optimization. This part is thus a natural candidate for parallelization to further improve performance.
\subsection{Parallelization}
\begin{figure}[h!]
    \centering
    \includegraphics[width=.65\linewidth]{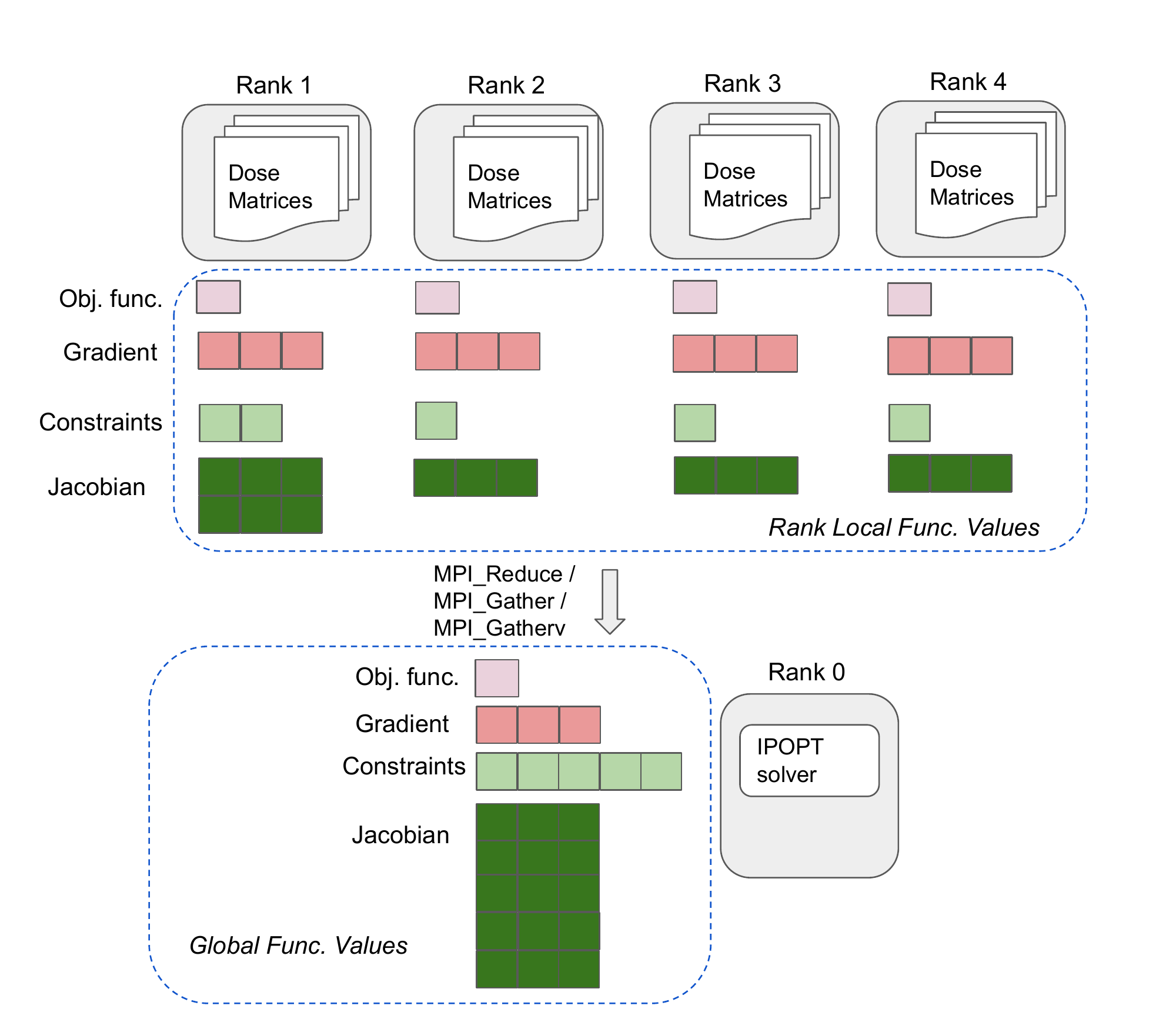}
    \caption{Illustration of the parallelization method used. Rank 0 holds the optimization solver and the global values for objective functions, constraints and gradients. The different terms and values are spread across the ranks, which compute local values that are aggregated back to Rank 0 using MPI collectives.}
    \label{fig:TROTS_parallelization}
\end{figure}

As mentioned in the previous section, function evaluation is a significant computational bottleneck in the optimization. This part thus becomes a natural target for parallelization, which can be achieved by distributing terms of the objective function and constraints between computational nodes. Indeed, that is the approach we use in this work.
\\ \\
We use MPI to distribute the computation, where MPI rank 0 (the leader) holds the IPOPT instance (which does not natively support MPI), and the remaining ranks compute objective function and constraints in parallel when requested by rank 0. The parallelization works such that each MPI process is assigned a set of terms of the objective function and constraints for which it is responsible for computing values. When rank 0 requires new function values, it broadcasts the current values of the optimization variables to all other processes, which then computes the function and gradient values for which it is responsible, before MPI collectives are used to aggregate the result to rank 0, which can then proceed with the next iteration in the optimization algorithm. A conceptual overview of the parallelization method is shown in Fig.~\ref{fig:TROTS_parallelization}.
\\ \\
Listing~\ref{mpi_pool} shows the code for the function handling the dispatching of function evaluations to the different MPI ranks which is the key in enabling the use of a distributed method for computing function values with an MPI-unaware optimization solver. At startup, when initialization is finished, all ranks, except rank 0, call the \texttt{compute\_vals\_mpi} function and wait at the \texttt{MPI\_Barrier}. When the optimization solver requires new function and constraint values to continue, it calls \texttt{compute\_vals\_mpi}, thus releasing the barrier and allowing all ranks to compute the required values in parallel.
\begin{lstlisting}[
    language=C++,
    keywordstyle=\color{blue}\ttfamily,
    stringstyle=\color{red}\ttfamily,
    commentstyle=\color{olive}\ttfamily,
    numbers=left,
    stepnumber=1,
    basicstyle=\tiny, %or \small or \footnotesize etc.
    caption={"Task-pool", function handling dispatching of function evaluations to different ranks. All ranks but 0 wait in the barrier at line 7 until rank 0 requests new function values to be computed.},
    captionpos=b,
    label=mpi_pool,
    columns=fullflexible,
    float
]
double compute_vals_mpi(bool calc_obj, const double* x, double* cons_vals,
                        bool calc_grad, double* grad,
                        LocalData& local_data,
                        std::optional<ConsDistributionData> distrib_data,
                        bool done) {
    while (true) {
        //"Task-pool", wait here until rank 0 is requesting function values to be computed
        MPI_Barrier(MPI_COMM_WORLD);

        //Check if rank 0 is signalling that the optimization is done
        int done_flag = static_cast<int>(done);
        MPI_Bcast(&done_flag, 1, MPI_INT, 0, MPI_COMM_WORLD);
        if (done_flag)
            return 0.0;

        int rank;
        MPI_Comm_rank(MPI_COMM_WORLD, &rank);
        double obj_val = 0.0;

        //Are we calculating objectives or constraints?
        int calc_obj_flag = static_cast<int>(calc_obj);
        MPI_Bcast(&calc_obj, 1, MPI_INT, 0, MPI_COMM_WORLD);
        if (calc_obj) {
            obj_val = compute_obj_vals_mpi(x, calc_grad, grad, local_data);
        } else {
            compute_cons_vals_mpi(x, cons_vals, calc_grad,grad, local_data, distrib_data);
        }

        //Rank 0 returns to the optimization solver
        //to continue to the next iteration / step
        if (rank == 0)
            return obj_val;
    }
}
\end{lstlisting}
Considering the uneven distribution in sizes of the dose matrices, as seen in Fig.~\ref{fig:nnz_hist}, a way to balance the workload between processes is required. When calculating objectives and constraints, the most computationally expensive part is the sparse matrix-vector product for the dose. Thus, we use the number of non-zeros in the dose influence matrices for each term to balance the workload between processors.
\\ \\
The problem of distributing the terms of the objectives and constraints as evenly as possible is an instance of the \emph{multi-way number partitioning problem} \cite{korf2009multi}, where one seeks a partitioning of a multiset of integers into $k$ partitions, such that the discrepancy between the sizes of the partitions is minimal. This problem is NP-complete, making heuristic algorithms attractive choices. A simple greedy heuristic is to sort the numbers in descending order, then, in order, assign the numbers to the partition with the smallest sum at that point. On average, one would expect this to give a partitioning with a discrepancy on the order of the smallest number \cite{korf2009multi}. Considering the difference in size between the smallest and largest dose influence matrices (again, see Fig.~\ref{fig:nnz_hist}), we expect the greedy heuristic to work well enough in our case. Thus, our method of distributing terms of the objectives and constraints is as follows:
\begin{enumerate}
    \item Sort the terms in descending order based on number of non-zeros in the corresponding dose matrix
    \item Go through the terms in order, assigning each term to the processor with the smallest total number of non-zeros.
\end{enumerate}
An example of a resulting distribution of matrices for the case with six total MPI ranks (recalling that one rank is reserved for the optimization solver) is shown in Fig.~\ref{fig:nnz_distrib}.
\subsection{Experimental Setup}
The performance experiments in this study are carried out on following systems:
\begin{itemize}
    \item \textbf{Dardel} is an HPE Cray EX supercomputer at PDC in Stockholm, Sweden. We use the \texttt{main} partition on Dardel where each node has two AMD EPYC 7742 CPUs.
    \item \textbf{Kebnekaise} is a supercomputer at HPC2N in Umeå, Sweden. Again we use the \texttt{Compute} partition of the cluster where each node has a single Intel Xeon E5-2690v4 CPU, 128 GB of RAM.
\end{itemize}
We compile our codes using GCC 11.2.0 on both systems. We refer to the Github repository of the code for the dependencies required to build our library. For MPI, we use Cray MPICH 8.1.11 on Dardel and OpenMPI 4.1.1 on Kebnekaise.
\section{Results}
\subsection{Performance and Parallel Scaling}
\begin{figure}[h!]
    \begin{subfigure}[b]{.47\textwidth}
        \centering
        \includegraphics[width=\textwidth]{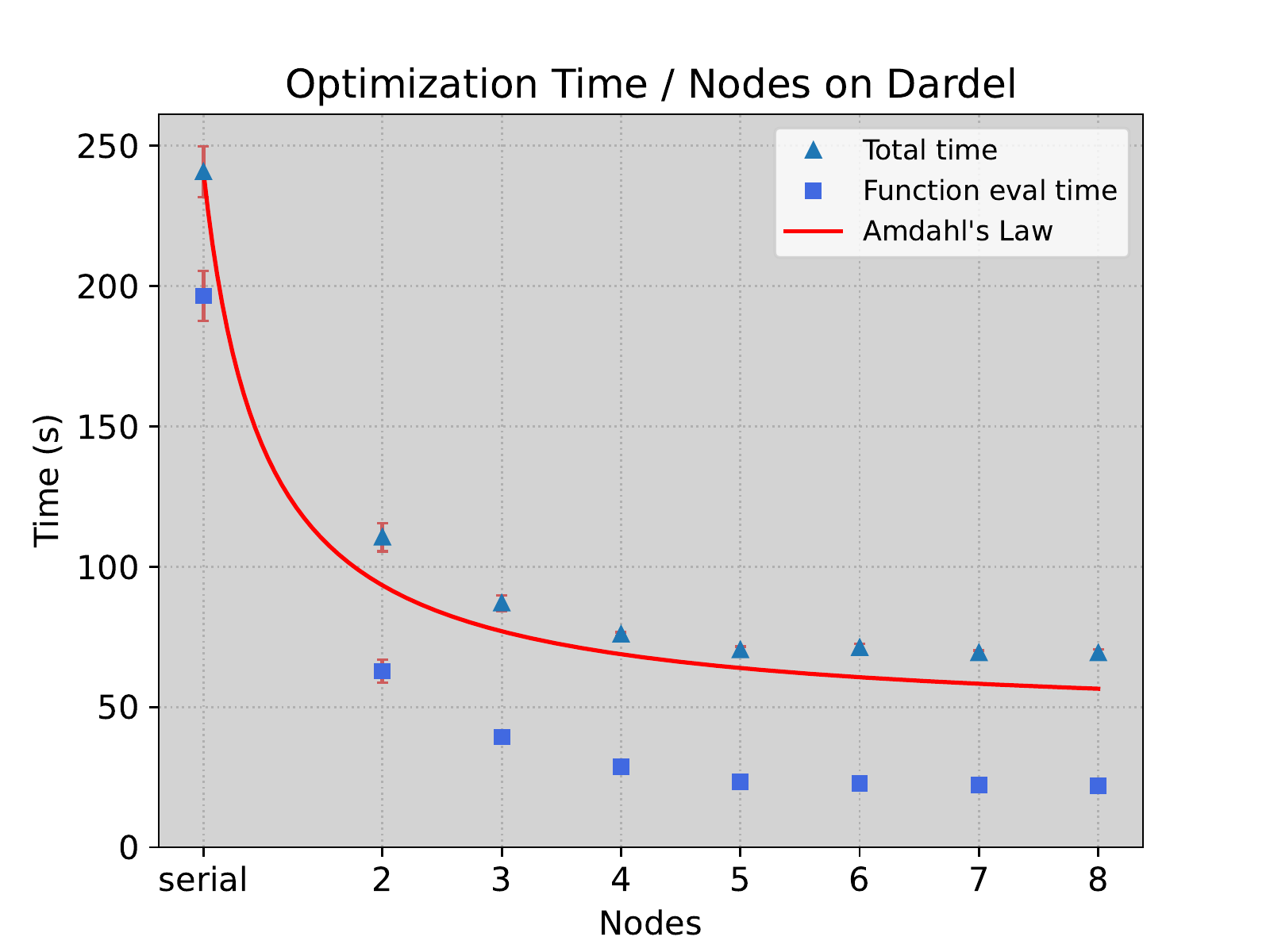}
        \caption{Parallel scaling on Dardel, the serial case uses half a node, since we only want to use one CPU socket to get a proper serial reference.}
        \label{fig:obj_nnz_distrib}
    \end{subfigure}
    \hfill
    \begin{subfigure}[b]{.47\textwidth}
        \centering
        \includegraphics[width=\textwidth]{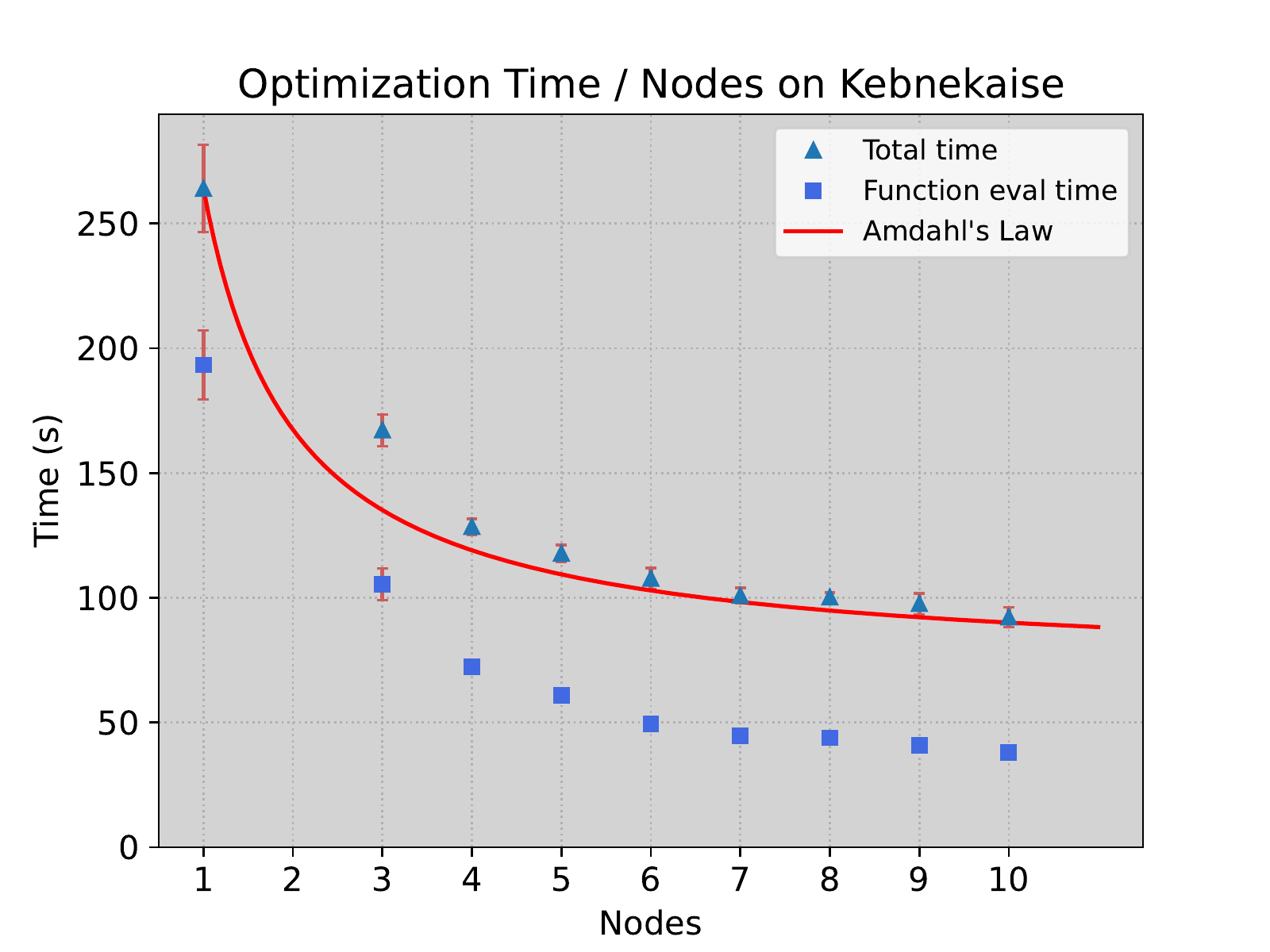}
        \caption{Parallel scaling on Kebnekaise. The 2 node case is omitted, since there is no parallelism in that case due to the optimization solver occupying one full node.}
        \label{fig:cons_nnz_distrib}
    \end{subfigure}
    \caption{Scaling tests on Dardel and Kebnekaise}
    \label{fig:Dardel_scaling}
\end{figure}
\noindent
We begin by assessing the performance and parallel scaling of our code on two supercomputing clusters.
Fig.~\ref{fig:Dardel_scaling} shows the total run time of the optimizer depending on the number of nodes. In all cases, the optimization was run for 3000 iterations, and repeated five times with the average times and standard deviations (vertical red lines) shown. All times were measured using IPOPT's internal timers, using the \texttt{print\_timing\_statistics} option. The upper data points (triangles) show the total execution time of the optimization. The lower data points (squares) show the portion of time in function evaluations only. The solid red line shows the theoretically possible time as predicted by Amdahl's law, where the serial portion is the time spent in IPOPT.
\subsection{Plan Quality}
\begin{figure}[h!]
    \centering
    \includegraphics[width=.85\linewidth]{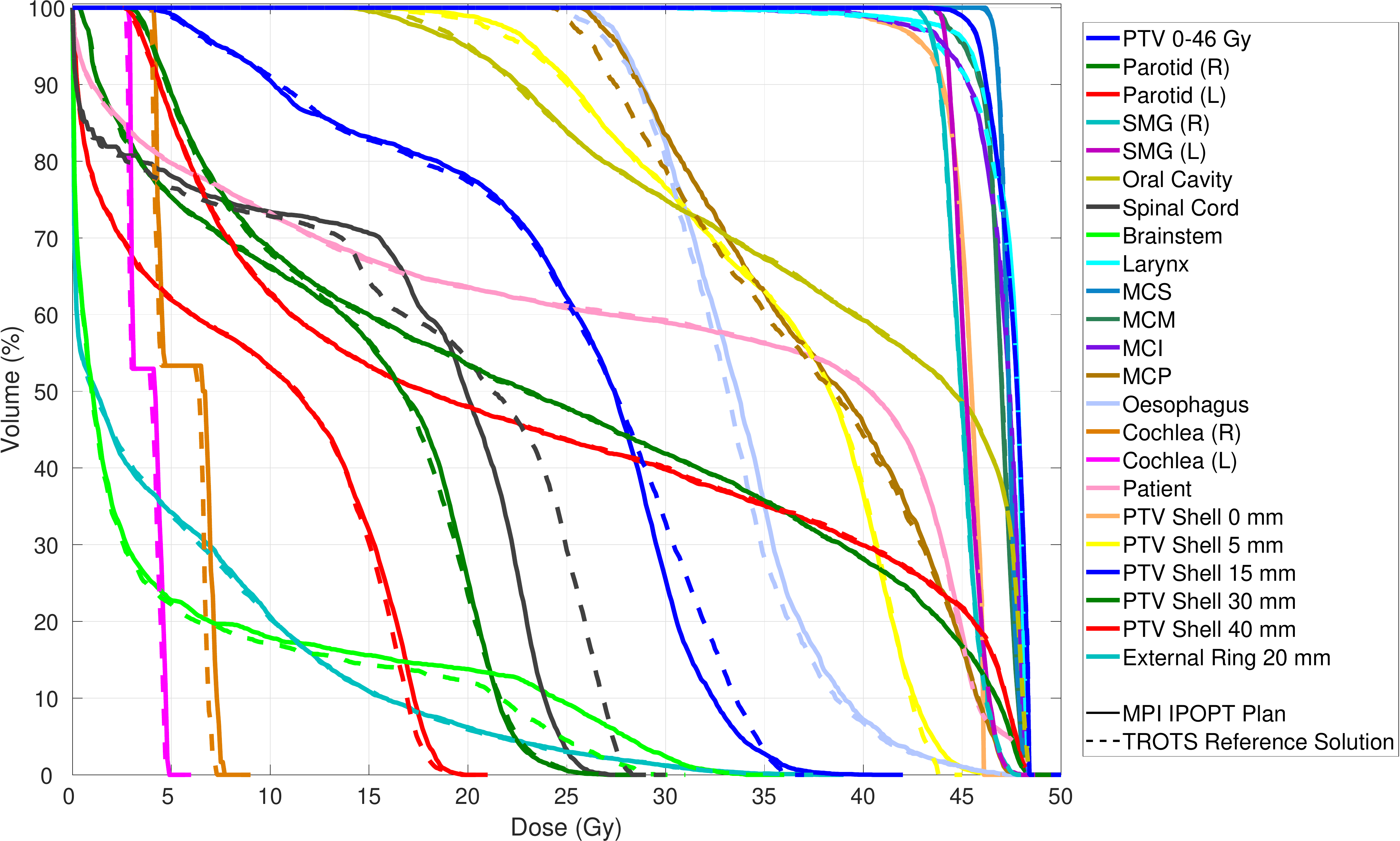}
    \caption[test]{DVH comparison between our plan, computed using IPOPT and 3 nodes on Dardel, and a reference plan from the TROTS dataset. The DVH curves from our plan are shown in solid lines and the reference plan is the dashed line. This plot was created using a slightly modified version of Matlab scripts provided by the TROTS authors\protect\footnotemark.}
    \label{fig:dardel_DVH}
\end{figure}
\noindent
To verify that our approach produces treatment plans of high quality, we compare DVH curves (see Section~\ref{sec:DVH}) from our plans with the reference solution provided from TROTS. Fig.~\ref{fig:dardel_DVH} shows DVH curves from our parallel implementation (solid lines) compared to a reference plan provided by TROTS (dashed lines) on the first VMAT Head-and-Neck case. While a complete discourse on evaluating plans based on DVH curves is out of scope for this paper, we can see that the DVH curves between our solution and the reference are quite similar. A general guideline for treatment plans is that the planning target volume (PTV), which encompasses the tumor, should receive a sufficiently high uniform dose, while organs at risk should receive as little dose as possible.
\footnotetext{\url{https://github.com/SebastiaanBreedveld/TROTS}}

\subsection{Performance Analysis and Execution Tracing}
\begin{figure}[h!]
    \centering
    \includegraphics[width=.95\linewidth]{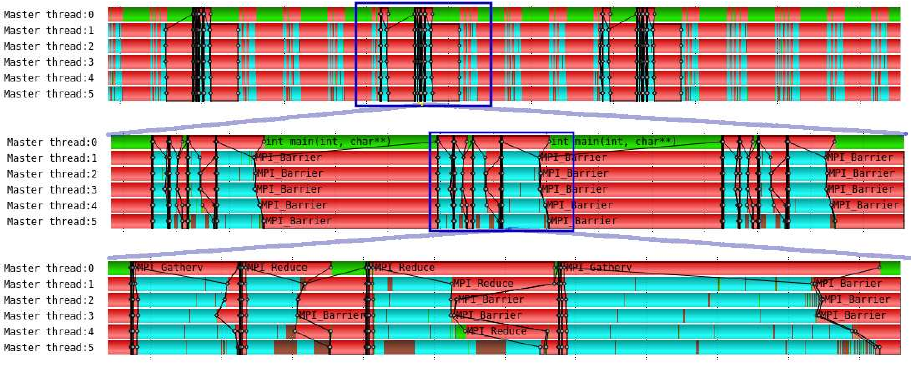}
    \caption{Tracing from a number of iterations of the optimization. Rank 0 handles the optimization solver (green is time spent in IPOPT) while the other ranks do the function evaluations where cyan is sparse linear algebra operations (to compute the dose), red is time waiting for the optimization solver, and the brown is computing function values from the dose. In the second panel we zoom in and show three evaluations. In the third panel the different stages of the objective function evaluation are clearly distinguishable.}
    \label{fig:traces}
\end{figure}
\begin{figure}[t]
    \centering
    \includegraphics[width=.9\linewidth]{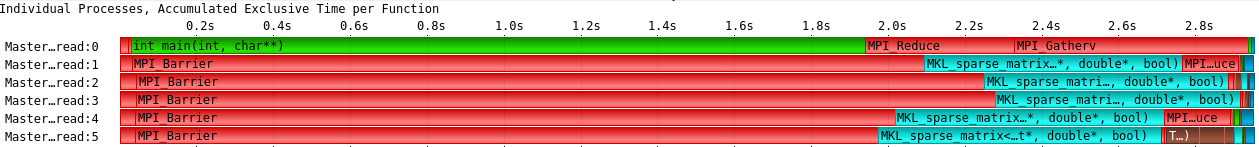}
    \caption{The accumulated exclusive time per function. The optimization solver is found on rank 0 and, green represents compute. The color scheme is the same as in Fig.~\ref{fig:traces}}
    \label{fig:acctime}
\end{figure}
\noindent
To further understand the performance of our code, we traced our applications using Score-P \cite{knupfer2012score}. We traced for a total of 100 iterations using 6 MPI ranks on the Dardel system, and the trace for a few iterations can be seen in Fig.~\ref{fig:traces}, which shows the tracing for a few iterations but zoomed in at different scales. The red bars show the idle time in \texttt{MPI\_Barrier} for each rank, the green bars in rank 0 is the time rank 0 spends in the IPOPT optimization solver. The cyan bars in the other ranks is the time computing the sparse matrix-vector products to evaluate function values, and the brown bars come from time spent in function evaluations outside the matrix-vector products.
\\ \\
From the tracing in Fig.~\ref{fig:traces} we see that our load balancing scheme works reasonably well, with the amount of time spent in the different ranks when computing function values being similar. There is some imbalance, especially when looking at rank 5, which appears to be caused by the function value evaluation from the dose. On closer investigation the function type causing the long evaluation times is the generalized mean, for which we use the C++ standard library function \texttt{std::pow} to compute the powers, which may be quite expensive. From the tracing we can also see the limitation in scaling imposed by the serial IPOPT solver, where the other MPI ranks are idle.
\\ \\
Fig.~\ref{fig:acctime} shows the accumulated time spent in different functions for the MPI ranks, with each row representing one MPI rank. The length of each bar shows the proportion of time spent in the corresponding function call. We see that the load balancing between the ranks is decent, but with some room for improvement.
\section{Discussion and Conclusion}
We have developed a parallel code for solving optimization problems from the TROTS dataset for radiation therapy treatment planning, capable of utilizing multi-node computational clusters for evaluating objective functions, constraints and their gradients. We show that our code can produce treatment plans of high quality while utilizing high-performance computing clusters effectively. Our approach distributes the function evaluations at each iteration across computational nodes, while using the state of the art single-node optimization solver IPOPT to compute the next iteration. We show that our approach can improve solution times by a factor of around 3.5 when compared to the serial time on a traditional supercomputer. While the possible parallel scaling is limited by the serial portion coming from the optimization solver, computational efficiency is often crucial for real clinics, and improvements in optimization times and time-to-solution are valued highly. 
    
%
%
%
\bibliographystyle{splncs04}
%
\bibliography{References}
\end{document}